\documentclass{article}
\usepackage{theorem,amsmath,amssymb}
\newtheorem{Th}{Theorem}
\newtheorem{Le}{Lemma}

\begin{document}
\title{A maximum modulus estimate for solutions of the Navier-Stokes system in domains of polyhedral type}
\date{}
\author{by V.~Maz'ya and J.~Rossmann}
\maketitle
\begin{abstract}
The authors prove a maximum modulus estimate for solutions of the
stationary Navier-Stokes system in bounded domains of polyhedral type. \\


\end{abstract}

\maketitle

\section{Introduction}

The present paper is concerned with solutions of the boundary value problem
\begin{equation} \label{a1}
-\nu\, \Delta v +  (v\cdot \nabla)\, v + \nabla p = 0,\quad \nabla \cdot v =0
   \ \mbox{ in } \Omega,\quad v|_{\partial\Omega}=\phi
\end{equation}
($\nu>0$), where $\Omega$ is a domain of polyhedral type. This
means that the boundary $\partial\Omega$ is the union of a finite
number of nonintersecting faces (two-dimensional open manifolds of
class $C^2$), edges (open arcs of class $C^2$), and vertices (the
endpoints of the edges). For every edge point or vertex $x_0$,
there exist a neighborhood $U$ and a diffeomorphism $\kappa: \,
U\to {\Bbb R}^3$ of class $C^2$ mapping $U\cap\Omega$ onto the
intersection of the unit ball with a polyhedron. Note that the
results of this paper are also valid for domains of the class
$\Lambda^2$ introduced in \cite{mp-83}.

It is well-known that the solution of the boundary value problem
\begin{equation} \label{b1}
-\Delta w + \nabla q = 0,\quad \nabla \cdot w =0 \ \mbox{ in }\Omega, \quad  w|_{\partial\Omega} = \phi
\end{equation}
for the linear Stokes system in a domain $\Omega\subset {\Bbb R}^3$ with smooth boundary $\partial\Omega$
satisfies the estimate
\begin{equation} \label{a3}
\| w\|_{L_\infty(\Omega)} \le c\, \| \phi\|_{L_\infty(\partial\Omega)}
\end{equation}
with a constant $c$ independent of $\phi$. This inequality was
first established without proof by Odquist \cite{Odquist}. A proof
of this inequality is given e.g. in the book by Ladyzhenskaya. We
refer also to the papers of Naumann \cite{Naumann} and Maremonti
\cite{Maremonti}. Using pointwise estimates of Green's matrix,
Maz'ya and Plamenevski\u{\i} \cite{mp-83} proved the inequality
(\ref{a3}) for solutions of problem (\ref{b1}) in domains of
polyhedral type.

For the nonlinear problem (\ref{a1}), Solonnikov \cite{Solonnikov}
showed that the solution satisfies the estimate
\begin{equation} \label{a2}
\| v\|_{L_\infty(\Omega)} \le c\, \big( \| \phi\|_{L_\infty(\partial\Omega)}\big),
\end{equation}
with a certain function $c$ if the boundary $\partial\Omega$ is
smooth. Maz'ya and Plamenevski\u{\i} \cite{mp-83} proved for
domains of polyhedral type that the solution $v$ of (\ref{a1})
with finite Dirichlet integral is continuous in $\bar{\Omega}$ if
$\phi$ is continuous on $\partial\Omega$. However, \cite{mp-83}
contains no estimates for the maximum modulus of $v$. The goal of
the present paper is to generalize Solonnikov's result to solutions of problem (\ref{a1})
in domains of polyhedral type and to obtain a more precise estimate. The
function $c$ constructed in the present paper has the form
\begin{equation} \label{a4}
c(t) = c_0t\, e^{c_1t/\nu},
\end{equation}
where $c_0$ and $c_1$ are positive constants independent of $\nu$.

\section{Estimates for solutions of the linear Stokes system}

First, we consider problem (\ref{b1}). Throughout this paper, we
assume that $\phi \in L_\infty(\partial\Omega)$ and
\begin{equation} \label{1bl1a}
\int_{\partial\Omega} \phi\cdot n\, d\sigma =0.
\end{equation}
The following two lemmas were proved in \cite{Solonnikov} for
domains with smooth boundaries. We give here other proofs which do
not require the smoothness of the boundary $\partial\Omega$. In
particular for the proof of Lemma 1, we will employ the estimates
of Green's matrix given in \cite{mp-83}.

\begin{Le} \label{bl1}
Let $\Omega$ be a domain of polyhedral type, and let $(w,q)$ be the solution of problem
{\em (\ref{b1})} satisfying the condition $\int_\Omega q(x)\, dx =0$. Then there exists
a constant $c$ independent of $\phi$ such that the inequalities {\em (\ref{a3})} and
\begin{equation}  \label{2bl1}
\sup_{x\in \Omega} d(x)\, \Big( \sum_{j=1}^3 \big| \partial_{x_j} w(x)\big| + \big|q(x)\big|\Big)
  \le   c\, \| \phi\|_{L_\infty(\partial\Omega)}
\end{equation}
are satisfied, where $d(x)=\mbox{\em dist}(x,\partial\Omega)$.
\end{Le}

P r o o f. As mentioned in the introduction, the inequality (\ref{a3}) was proved in
\cite[Cor.9.2]{mp-83}. We include its proof for readers' convenience. Let
$G(x,\xi) = \big( G_{i,j}(x,\xi)\big)_{i,j=1}^4$ denote the Green matrix for problem
(\ref{b1}). This means that the vectors $\vec{G}_j=(G_{1,j},G_{2,j},G_{3,j})$ and the function
$G_{4,j}$ are the uniquely determined solutions of the problems
\begin{eqnarray*}
&& -\Delta_x \vec{G}_j(x,\xi)+ \nabla_x G_{4,j}(x,\xi) = \delta(x-\xi)\, \vec{e}_j,\quad
  \nabla_x\cdot \vec{G}_j(x,\xi) = 0 \quad\mbox{ for }x,\xi \in \Omega, \ j=1,2,3,\\
&& -\Delta_x \vec{G}_4(x,\xi)+ \nabla_x G_{4,4}(x,\xi) = 0,\quad
  \nabla_x\cdot \vec{G}_4(x,\xi) = \delta(x-\xi)- (\mbox{mes}(\Omega))^{-1} \quad\mbox{ for }x,\xi \in \Omega, \\
&& \vec{G}_j(x,\xi)=0 \ \mbox{ for }x\in \partial\Omega,\ \xi \in
\Omega,\ j=1,2,3,4,
\end{eqnarray*}
satisfying the condition
\[
\int_\Omega G_{4,j}(x,\xi)\, dx =0 \mbox{ for }\xi \in \Omega,\ j=1,2,3,4.
\]
Here $\vec{e}_j$ denotes the vector $(\delta_{1,j},\delta_{2,j},\delta_{3,j})$. Then the components of
the vector function $w$ and  $q$ have the representation
\begin{eqnarray*}
w_i(x) & = & \int\limits_{\partial\Omega} \Big( -\sum_{j=1}^3
  \frac{\partial G_{i,j}(x,\xi)}{\partial n_\xi}\,
  \phi_j(\xi) + G_{i,4}(x,\xi)\, \phi(\xi)\cdot n_\xi \Big)\, d\sigma_\xi, \ i=1,2,3,\\
q(x) & = & \int\limits_{\partial\Omega} \Big( -\sum_{j=1}^3
  \frac{\partial G_{4,j}(x,\xi)}{\partial n_\xi}\,
  \phi_j(\xi) + G_{4,4}(x,\xi)\, \phi(\xi)\cdot n_\xi \Big)\, d\sigma_\xi.
\end{eqnarray*}
For the proof of (\ref{a3}) and (\ref{2bl1}), we employ the estimates of the functions $G_{i,j}$
given in \cite{mp-83} (for a more general boundary value problem in a cone with edges see also
\cite{mr-05}). We start with the inequality (\ref{2bl1}).
Suppose that $x$ lies in a neighborhood ${\cal U}$ of the vertex $x^{(1)}$.
We denote by ${\cal S}$ the set of the vertices and edge points of the boundary $\partial\Omega$, by
$\rho_i(x)$ the distance of $x$ from the vertex $x^{(i)}$, by
$r_k(x)$ the distance from the edge $M_k$, by $r(x)=\min_k r_k(x)$
the distance from the set of all edge points, and introduce the
following subsets of ${\cal U}\cap (\partial\Omega \backslash {\cal S})$:
\begin{eqnarray*}
&& E_1 = \{ \xi \in {\cal U}\cap (\partial\Omega \backslash {\cal S}): \ \rho_1(\xi)>2\rho_1(x) \}, \\
&& E_2 = \{ \xi \in {\cal U}\cap (\partial\Omega \backslash {\cal S}): \ \rho_1(\xi)<\rho_1(x)/2 \}, \\
&& E_3 = \{ \xi \in {\cal U}\cap (\partial\Omega \backslash {\cal S}): \
  \rho_1(x)/2 < \rho_1(\xi)<2\rho_1(x), \ |x-\xi|>\min(r(x),r(\xi)) \}, \\
&& E_4 = \{ \xi \in {\cal U}\cap (\partial\Omega \backslash {\cal S}): \
  \rho_1(x)/2 < \rho_1(\xi)<2\rho_1(x), \ |x-\xi|<\min(r(x),r(\xi)) \}.
\end{eqnarray*}
Let $K(x,\xi)$ be one of the functions
\[
\frac{\partial}{\partial x_j} \, \frac{\partial}{\partial n_\xi}\, G_{i,j}(x,\xi), \quad
  \frac{\partial G_{i,4}(x,\xi)}{\partial x_j}\, , \quad
  \frac{\partial}{\partial n_\xi} G_{4,j}(x,\xi), \quad G_{4,4}(x,\xi),
\]
$i,j=1,2,3$. Then the following estimates are valid for $x\in {\cal U},\ \xi \in {\cal U}
\cap (\partial\Omega \backslash {\cal S})$:
\begin{eqnarray*}
\big| K(x,\xi)\big| & \le & c\, \rho_1(x)^{\Lambda-1}\,
  \rho_1(\xi)^{-\Lambda-2}  \prod_{k \in J_1} \Big( \frac{r_k(x)}{\rho_1(x)}\Big)^{\mu_k-1} \, \prod_{k\in J_1}
  \Big( \frac{r_k(\xi)}{\rho_1(\xi)}\Big)^{\mu_k-1} \quad\mbox{for }\xi \in E_1, \\
\big| K(x,\xi)\big| & \le & c\, \rho_1(x)^{-\Lambda-2}\, \rho_1(\xi)^{\Lambda-1}
  \prod_{k \in J_1} \Big( \frac{r_k(x)}{\rho_1(x)}\Big)^{\mu_k-1} \, \prod_{k\in J_1}
  \Big( \frac{r_k(\xi)}{\rho_1(\xi)}\Big)^{\mu_k-1} \quad\mbox{for }\xi \in E_2, \\
\big| K(x,\xi)\big| & \le & c\, |x-\xi|^{-3} \, \Big(
\frac{r(x)}{|x-\xi|}\Big)^{\mu-1} \  \Big( \frac{r(\xi)}{|x-\xi|}\Big)^{\mu-1} \quad\mbox{for }\xi \in E_3, \\
\big| K(x,\xi)\big| & \le & c\,  |x-\xi|^{-3} \quad\mbox{for }\xi \in E_4,
\end{eqnarray*}
where $\Lambda>0$, $\mu_k>1/2$, $\mu >1/2$. Here $J_l$ is the set
of all indices $k$ such that $x^{(l)}\in \overline{M}_k$. Note that
\[
c_1\, r(x)\le \rho_1(x) \,  \prod_{k \in J_1} \frac{r_k(x)}{\rho_1(x)} \le c_2\, r(x) \quad\mbox{for }
  x\in {\cal U},
\]
where $c_1$ and $c_2$ are positive constants. We consider the
integral
\[
I(x) = \int\limits_{\partial\Omega\cap {\cal U}} K(x,\xi)\, \psi(\xi)\, d\sigma_\xi
\]
for $x\in {\cal U}$, $\psi \in L_\infty(\partial\Omega)$ and write
this integral as a sum $I(x)=I_1+I_2+I_3+I_4$, where $I_k$ is the
integral of $K(x,\xi)\, \psi(\xi)$ over the set $E_k$,
$k=1,2,3,4$. Then
\begin{eqnarray*}
I_1 & \le & c\, \rho_1(x)^{\Lambda-1} \,  \prod_{k \in J_1} \Big(
\frac{r_k(x)}{\rho_1(x)}\Big)^{\mu_k-1}
  \, \| \psi\|_{L_\infty(\partial\Omega)} \int_{E_1} \rho_1(\xi)^{-\Lambda-2}  \prod_{k \in J_1}\,
  \Big( \frac{r_k(\xi)}{\rho_1(\xi)}\Big)^{\mu_k-1}\, d\sigma_\xi\\
& \le & c\, \rho_1(x)^{-1} \,  \prod_{k \in J_1} \Big(
\frac{r_k(x)}{\rho_1(x)}\Big)^{\mu_k-1}
  \, \| \psi\|_{L_\infty(\partial\Omega)} \le c\, r(x)^{-1}  \, \| \psi\|_{L_\infty(\partial\Omega)}\, .
\end{eqnarray*}
Analogously, the inequality
\[
I_2 \le c\, r(x)^{-1}  \, \| \psi\|_{L_\infty(\partial\Omega)}
\]
holds. Suppose without loss of generality that $M_1$ is the nearest edge to $x$. We denote by
$E_3^{(1)}$ the set of all $\xi\in E_3$ such that $r(\xi)< r_1(\xi)$. Furthermore, let
$I_3^{(1)}$ be the integral of $K(x,\xi)\, \psi(\xi)$ over the set $E_3^{(1)}$. If $\xi \in E_3^{(1)}$,
then there exists a positive constant $c$ such that $|x-\xi|> c\, \rho_1(x)$. Hence
\[
I_3^{(1)} \le c\, \rho_1(x)^{-2\mu-1}\, r_1(x)^{\mu-1} \, \|
  \psi\|_{L_\infty(\partial\Omega)} \,   \int_{E_3^{(1)}} r(\xi)^{\mu-1}\, d\sigma_\xi .
\]
Since $E_3^{(1)}\subset\{ \xi:\, \rho_1(x)/2 < \rho_1(\xi) <
2\rho_1(x)\}$ and $r_1(x)\le\rho_1(x)$, we obtain
\[
I_3^{(1)} \le c\, \rho_1(x)^{-\mu} r_1(x)^{\mu-1} \, \| \psi\|_{L_\infty(\partial\Omega)} \le
  c\, r_1(x)^{-1} \, \| \psi\|_{L_\infty(\partial\Omega)}\, .
\]
Let $\xi \in E_3\backslash E_3^{(1)}$ and let $x'$, $\xi'$ denote the nearest points on the
edge $M_1$ to $x$ and $\xi$, respectively. Then there exists a positive constant $c$
independent of $x$ and $\xi$ such that
\[
|x-\xi| > c\, \big( r(x)+r(\xi)+|x'-\xi'|\big).
\]
Consequently,
\begin{eqnarray*}
| I_3 -I_3^{(1)} | & \le & c\, r(x)^{\mu-1}\, \| \psi\|_{L_\infty(\partial\Omega)}\,
  \int\limits_{E_3\backslash E_3^{(1)}}   \frac{r(\xi)^{\mu-1}}{(r(x)+r(\xi)+|x'-\xi'|)^{2\mu+1}}\, d\sigma_\xi \\
& \le & c\, r(x)^{\mu-1} \, \| \psi\|_{L_\infty(\partial\Omega)}\,
\int_0^\infty \int_{\Bbb R} \frac{r^{\mu-1}}{(r+r(x)+|t|)^{2\mu+1}}\, d\xi'\, dr
  = C\, r(x)^{-1}\, \| \psi\|_{L_\infty(\partial\Omega)} \, .
\end{eqnarray*}
Finally using the estimate for $K(x,\xi)$ in $E_4$, we obtain
\[
I_4 \le c\, \| \psi\|_{L_\infty(\partial\Omega)}\, \int_{E_4} |x-\xi|^{-3}\, d\sigma_\xi
  \le C \, d(x)^{-1}  \| \psi\|_{L_\infty(\partial\Omega)}\, .
\]
Thus we have shown that
\[
I(x) \le c\, d(x)^{-1} \, \| \psi\|_{L_\infty(\partial\Omega)}\quad \mbox{for }x\in \Omega\cap
{\cal U}.
\]
Now, we consider the integral
\begin{equation} \label{Iofx}
\int\limits_{\partial\Omega\cap {\cal V}} K(x,\xi)\, \psi(\xi)\, d\sigma_\xi
\end{equation}
for $x\in \Omega\cap {\cal U}$, where ${\cal V}$ is a neighborhood
of the vertex $x^{(l)}$, $\l\not=1$. Using the estimate
\[
\big| K(x,\xi)\big|  \le  c\, \rho_1(x)^{\Lambda-1}\,
  \rho_l(\xi)^{\Lambda-1}  \prod_{k \in J_1} \Big( \frac{r_k(x)}{\rho_1(x)}\Big)^{\mu_k-1} \, \prod_{k\in J_l}
  \Big( \frac{r_k(\xi)}{\rho_l(\xi)}\Big)^{\mu_k-1} \quad\mbox{for } x \in {\cal U}, \ \xi \in {\cal V},
\]
we obtain
\[
\Big| \int\limits_{\partial\Omega\cap {\cal V}} K(x,\xi)\, \psi(\xi)\, d\sigma_\xi \Big|
  \le c \,  \rho_1(x)^{\Lambda-1}\, \prod_{k \in J_1} \Big( \frac{r_k(x)}{\rho_1(x)}\Big)^{\mu_k-1}\,
  \| \psi\|_{L_\infty(\partial\Omega)} \le c\, r(x)^{-1} \, \| \psi\|_{L_\infty(\partial\Omega)}\, .
\]
The same estimate holds for the integral (\ref{Iofx}) in the case
when ${\cal V}$ is a neighborhood of an arbitrary other boundary
point. This proves (\ref{2bl1}). Analogously, (\ref{a3}) holds by
means of the estimates
\begin{eqnarray*}
\big| K(x,\xi)\big| & \le & c\, \rho_1(x)^{\Lambda}\,
  \rho_1(\xi)^{-\Lambda-2}  \prod_{k \in J_1} \Big( \frac{r_k(x)}{\rho_1(x)}\Big)^{\mu_k} \, \prod_{k\in J_1}
  \Big( \frac{r_k(\xi)}{\rho_1(\xi)}\Big)^{\mu_k-1} \quad\mbox{for }\xi \in E_1, \\
\big| K(x,\xi)\big| & \le & c\, \rho_1(x)^{-\Lambda-1}\,
\rho_1(\xi)^{\Lambda-1}
  \prod_{k \in J_1} \Big( \frac{r_k(x)}{\rho_1(x)}\Big)^{\mu_k} \, \prod_{k\in J_1}
  \Big( \frac{r_k(\xi)}{\rho_1(\xi)}\Big)^{\mu_k-1} \quad\mbox{for }\xi \in E_2, \\
\big| K(x,\xi)\big| & \le & c\, |x-\xi|^{-2} \, \Big(
\frac{r(x)}{|x-\xi|}\Big)^{\mu} \  \Big( \frac{r(\xi)}{|x-\xi|}\Big)^{\mu-1} \quad\mbox{for }\xi \in E_3, \\
\big| K(x,\xi)\big| & \le & c\,  d(x)\, |x-\xi|^{-3}
\quad\mbox{for }\xi \in E_4,
\end{eqnarray*}
for the functions $K(x,\xi)=\partial G_{i,j}(x,\xi)/\partial
n_\xi$ and $K(x,\xi)=G_{i,4}(x,\xi)$,
$i,j=1,2,3$ (see \cite[Th.9.1]{mp-83}). \hfill $\Box$ \\

We denote by $W^{l,p}(\Omega)$ the Sobolev space with the norm
\[
\| u \|_{W^{l,p}(\Omega)} = \Big( \int_\Omega \sum_{|\alpha|\le l}
\big| \partial_x^\alpha u(x)\big|^p\, dx\Big)^{1/p}.
\]
Here $l$ is a nonnegative integer and $1<p<\infty$.

\begin{Le} \label{bl3}
Let $(w,q)$ be a solution of problem {\em (\ref{b1})}, where  $\Omega$ is a domain of polyhedral type.
Then there exists a vector function $b \in W^{1,6}(\Omega)^3$ such that $w=\mbox{\em rot}\, b$ and
\begin{equation} \label{1bl3}
\| b \|_{W^{1,6}(\Omega)} \le c\, \| \phi\|_{L_\infty(\partial\Omega)}
\end{equation}
with a constant $c$ independent of $\phi$.
\end{Le}

P r o o f. Let $B_\rho$ be a ball with radius $\rho$ centered at the origin and such that
$\overline\Omega \subset B_\rho$. Furthermore, let $(w^{(1)},s)$ be a solution of the problem
\[
-\Delta w^{(1)} + \nabla s = 0,\quad \nabla \cdot w^{(1)} =0 \
\mbox{in }B_\rho\backslash \overline\Omega, \quad
  w^{(1)}|_{\partial\Omega} = \phi, \quad w^{(1)}|_{\partial B_\rho}=0.
\]
Obviously, the vector function
\[
u(x) = \left\{ \begin{array}{ll} w(x) & \mbox{for } x\in \Omega,\\
  w^{(1)}(x) & \mbox{for }x\in B_\rho\backslash \Omega \end{array} \right.
\]
satisfies the equality $\nabla\cdot u=0$ in the sense of
distributions in $B_\rho$. Due to Lemma \ref{bl1}, the $L_\infty$
norms of $w$ and $w^{(1)}$ can be estimated by the $L_\infty$ norm
of $\phi$. Hence,
\[
\| u\|_{L_6(B_\rho)} \le c\, \| \phi
\|_{L_\infty(\partial\Omega)}\, ,
\]
where $c$ is a constant independent of $\phi$. Suppose that there
exists a vector function $U\in W^{2,6}(B_\rho)^3$ satisfying the
equations
\begin{equation} \label{2dbl3}
-\Delta U = u \ \mbox{ in }B_\rho,\quad \nabla \cdot U = 0\ \mbox{
on }\partial B_\rho
\end{equation}
and the inequality
\begin{equation} \label{3dbl3}
\| U\|_{W^{2,6}(B_\rho)^3} \le c\, \| u\|_{L_6(B_\rho)^3}\, .
\end{equation}
Since $\Delta(\nabla\cdot U)= \nabla\cdot u =0$ in $B_\rho$ it
follows that $\nabla\cdot U =0$ in $B_\rho$. Consequently for the
vector function $b=\mbox{rot}\, U$, we obtain
\[
\mbox{rot}\, b = \mbox{rot\,rot}\, U = -\Delta U +
\mbox{grad\,div}\, U = u \ \mbox{ in }B_\rho
\]
and
\[
\| b\|_{W^{1,6}(B_\rho)^3} \le c_1\, \| U\|_{W^{2,6}(B_\rho)^3}
\le c\, c_1\, \| u\|_{L_6(B_\rho)^3}
  \le c_2\, \| \phi \|_{L_\infty(\partial\Omega)}\, .
\]
It remains to show that problem (\ref{2dbl3}) has a solution $U$
subject to (\ref{3dbl3}). To this end, we consider the boundary
value problem
\begin{equation} \label{1bl2}
- \Delta U = u \ \mbox{ in } B_\rho,\quad \frac{\partial
U_r}{\partial r} + \frac 2r \, U_r = U_\theta = U_\varphi
  = 0 \ \mbox{ on }\partial B_\rho,
\end{equation}
where $U_r,U_\theta,U_\varphi$ are the spherical components of the
vector function $U$, i.e.
\[
\left( \begin{array}{c} U_r \\ U_\theta \\ U_\varphi
\end{array}\right) =
\left( \begin{array}{ccc} \sin\theta\, \cos\varphi & \sin\theta\, \sin\varphi & \cos\theta \\
  \cos\theta\, \cos\varphi & \cos\theta\, \sin\varphi & -\sin\theta \\ -\sin\varphi &
  \cos\varphi & 0 \end{array}\right)\ \left( \begin{array}{c} U_1 \\ U_2 \\ U_3 \end{array}\right)\, .
\]
On the set of all $U$ satisfying the boundary conditions in
(\ref{1bl2}), we have
\begin{eqnarray*}
&& - \int_{B_\rho} \Delta U\cdot \bar{U}\, dx  =  \sum_{j=1}^3
\int_{B_\rho} \big| \partial_{x_j} U\big|^2\, dx
  - \rho^{-1}\, \int_{\partial B_\rho} \frac{\partial U}{\partial r} \cdot \bar{U}\, d\sigma  \\
&&  =  \sum_{j=1}^3 \int_{B_\rho} \big| \partial_{x_j} U\big|^2\,
dx
  - \rho^{-1} \, \int_{\partial B_\rho} \frac{\partial U_r}{\partial r} \cdot \bar{U}_r\, d\sigma
  = \sum_{j=1}^3 \int_{B_\rho} \big| \partial_{x_j} U\big|^2\, dx
   + 2\rho^{-2} \int_{\partial B_\rho} |U_r|^2\, d\sigma .
\end{eqnarray*}
Since the quadratic form on the right-hand side is coercive,
problem (\ref{1bl2}) is uniquely solvable in $W^{1,2}(B_\rho)^3$.
By a well-known regularity result for solutions of elliptic
boundary value problems, the solution belongs to
$W^{2,6}(B_\rho)^3$ and satisfies (\ref{3dbl3}) if $u \in
L_6(B_\rho)^3$. From (\ref{1bl2}) and from the equality
\[
\nabla\cdot U = \frac{\partial U_r}{\partial r} + \frac 2r \, U_r
+ \frac 1r \frac{\partial U_\theta}{\partial\theta}
  + \frac{\cot \theta}r\, U_\theta + \frac{1}{r\sin\theta}\, \frac{\partial U_\varphi}{\partial\varphi}
\]
it follows that $\nabla\cdot U =0$ on $\partial B_\rho$. The proof of the lemma is complete. \hfill $\Box$\\

Next, we consider the solution $(W,Q)$ of the problem
\begin{equation} \label{b2}
-\Delta W + \nabla Q = f,\quad \nabla \cdot W =0 \ \mbox{ in
}\Omega, \quad
 W|_{\partial\Omega} = 0 .
\end{equation}
Suppose that $x^{(1)},\ldots,x^{(d)}$ are the vertices and
$M_1,\ldots,M_m$ the edges of $\Omega$. As in the proof of Lemma
\ref{bl1}, we use the notation $\rho_j(x)=\mbox{dist}(x,x^{(j)})$,
$r_k(x)=\mbox{dist}(x,M_k)$, $\rho(x) = \min_j \rho_j(x)$, and
$r(x)=\min_k r_k(x)$. Then $V_{\beta,\delta}^{l,s}(\Omega)$ is
defined as the weighted Sobolev space with the norm
\[
\| u\|_{V_{\beta,\delta}^{l,s}(\Omega)} = \Big( \int_\Omega
\sum_{|\alpha|\le l} r(x)^{s(|\alpha|-m)}
  \prod_{j=1}^d \rho_j^{s\beta_j} \ \prod_{k=1}^m \Big(\frac{r_k}{\rho}\Big)^{s\delta_k}\ \big|
  \partial_x^\alpha u(x)\big|^s\, dx\Big)^{1/s}.
\]
Here, $l$ is a nonnegative integer, $s\in (1,\infty)$,
$\beta=(\beta_1,\ldots,\beta_d)\in {\Bbb R}^d$, and
$\delta=(\delta_1,\ldots,\delta_m)\in {\Bbb R}^m$. The space
$V_{\beta,\delta}^{-1,s}(\Omega)$ is the set of all distributions
of the form $u=u_0 +\nabla\cdot u^{(1)}$, where $u_0 \in
V_{\beta+1,\delta+1}^{0,s}(\Omega)$ and $u^{(1)} \in
V_{\beta,\delta}^{0,s}(\Omega)^3$. By Theorem \cite[Th.6.1]{mp-83}
(for a more general boundary value problem see also
\cite{mr-04b}), problem (\ref{b2}) is uniquely solvable (up to
vector functions of the form $(0,c)$, where $c$ is a constant) in
$V_{\beta,\delta}^{1,s}(\Omega)^3 \times
V_{\beta,\delta}^{0,s}(\Omega)$ for arbitrary $f\in
V_{\beta,\delta}^{-1,s}(\Omega)^3$ if
\[
|\beta_j- 3/2 + 3/s|< \varepsilon_j + 1/2 \quad \mbox{and}\quad
|\delta_k-1+2/s|< \varepsilon'_k +1/2 \, .
\]
Here $\varepsilon_j$ and $\varepsilon'_k$ are positive numbers
depending on $\Omega$. In particular, problem (\ref{b2}) has a
unique (up to constant $Q$) solution $(W,Q) \in
V_{0,0}^{1,s}(\Omega)^3 \times V_{0,0}^{0,s}(\Omega)$  satisfying
the estimate
\begin{equation} \label{b3}
\| W\|_{ V_{0,0}^{1,s}(\Omega)}  \le   c\, \| f\|_{
V_{0,0}^{1,s}(\Omega)}
\end{equation}
for arbitrary $f\in V_{0,0}^{-1,s}(\Omega)^3$ if $1<s<
3+\varepsilon$ with a certain $\varepsilon>0$. The components of
the vector function $W$ admit the representation
\begin{equation} \label{b4}
W_i(x) = \int_\Omega \sum_{j=1}^3 G_{i,j}(x,\xi)\, f_j(\xi)\,
d\xi,
\end{equation}
where $G_{i,j}(x,\xi)$ are the elements of Green's matrix
introduced in the proof of Lemma \ref{bl1}. From (\ref{b3}), we
obtain the following estimates.

\begin{Le} \label{bl4}
Suppose that $f=\partial_{x_j}g$, where $j\in \{1,2,3\}$. If $g\in
L_s(\Omega)^3$, $s>3$, then
\begin{equation} \label{1bl4}
\| W\|_{L_\infty(\Omega)} \le c\, \| g\|_{L_s(\Omega)}\, .
\end{equation}
If $g\in L_3(\Omega)^3$, then
\begin{equation} \label{2bl4}
\| W\|_{L_s(\Omega)} \le c\, \| g\|_{L_3(\Omega)}
\end{equation}
for arbitrary $s$, $1<s<\infty$.
\end{Le}

P r o o f. Let $g\in L_s(\Omega)$, $s>3$, and let $\varepsilon$ be
a sufficiently small positive number, $\varepsilon<s-3$. Then it
follows from (\ref{b3}) and from the continuity of the imbeddings
$V_{0,0}^{1,3+\varepsilon}(\Omega) \subset
W^{1,3+\varepsilon}(\Omega) \subset L_\infty(\Omega)$ that
\[
\| W\|_{L_\infty(\Omega)} \le c_1\, \|
W\|_{W^{1,3+\varepsilon}(\Omega)} \le c_2\,
  \| W\|_{V_{0,0}^{1,3+\varepsilon}(\Omega)} \le c_3\, \| g \|_{L_{3+\varepsilon}(\Omega)}
  \le c_4\, \| g \|_{L_s(\Omega)} \, .
\]
Analogously, we obtain
\[
\| W\|_{L_s(\Omega)} \le c_5\, \| W\|_{W^{1,3}(\Omega)} \le c_6\,
  \| W\|_{V_{0,0}^{1,3}(\Omega)} \le c_7\, \|g\|_{L_3(\Omega)} \, .
\]
The lemma is proved. \hfill $\Box$

\section{An estimate of the maximum modulus of the solution to the Navier-Stokes system}

Now we prove the main result of this paper introducing some modifications into Solonnikov's
scheme.

\begin{Th} \label{ct1}
Let $(v,q)$ be a solution of problem {\em (\ref{a1})}, where
$\Omega$ is a domain of polyhedral type. Then $v$ satisfies the
estimate {\em (\ref{a2})} with a function $c$ of the form {\em
(\ref{a4})}.
\end{Th}

P r o o f. Suppose first that $\nu=1$. Let $(w,q)$ be the solution
of problem (\ref{b1}), $\int_\Omega q(x)\, dx =0$. Then the vector
function $(v-w,p-q)$ satisfies the equations
\[
-\Delta (v-w) + \nabla (p-q) = - (v\cdot\nabla)\, v,\quad \nabla \cdot (v-w)=0
\]
in $\Omega$ and the boundary condition $v-w=0$ on
$\partial\Omega$. Hence by (\ref{b4}), we have $v=w+W$, where $W$
is the vector function with the components
\[
W_i(x) = - \int_\Omega \sum_{j=1}^3 G_{i,j}(x,\xi) \, \big(
v(\xi)\cdot \nabla\big)\, v_j(\xi)\, d\xi
   =  - \int_\Omega \sum_{j=1}^3 G_{i,j}(x,\xi) \, \nabla \cdot\big( v_j(\xi)\, v(\xi)\big)\, d\xi,
\]
$i=1,2,3$. Using (\ref{1bl4}), we obtain
\begin{eqnarray}
\| v\|_{L_\infty(\Omega)} & \le & \| w\|_{L_\infty(\Omega)} + \|
W\|_{L_\infty(\Omega)}
  \le \| w\|_{L_\infty(\Omega)} + c \, \sum_{i,j=1}^3 \| v_i\, v_j\|_{L_{s/2}(\Omega)} \nonumber \\ \label{1ct1}
& \le & \| w\|_{L_\infty(\Omega)} + c \, \| v\|^2_{L_s(\Omega)}
\end{eqnarray}
for arbitrary $s>6$. From (\ref{2bl4}) it follows that
\begin{eqnarray}
\| v\|_{L_s(\Omega)} & \le & \| w\|_{L_s(\Omega)} + \|
W\|_{L_s(\Omega)}
   \le   \| w\|_{L_s(\Omega)} +c \, \sum_{i,j=1}^3 \| v_i\, v_j\|_{L_3(\Omega)} \nonumber \\ \label{2ct1}
& \le & c_1\, \| w\|_{L_\infty(\Omega)} + c_2 \, \|
v\|^2_{L_6(\Omega)} \, .
\end{eqnarray}
Combining (\ref{a3}), (\ref{1ct1}) and (\ref{2ct1}), we obtain
\begin{equation} \label{4ct1}
\| v\|_{L_\infty(\Omega)} \le c_3\, \Big( \|
\phi\|_{L_\infty(\partial\Omega)}
  +  \| \phi\|^2_{L_\infty(\partial\Omega)} +  \| v\|^4_{L_6(\Omega)}\Big) .
\end{equation}
with a certain constant $c_3$ independent of $\phi$.

The norm of $v$ in $L_6(\Omega)$ can be estimated in the same way
as in \cite{Solonnikov}. Let $\delta(x)$ be the regularized distance of $x$ from the
boundary $\partial\Omega$ (see \cite[Ch.6,\S 2]{Stein}), i.e.
$\delta$ is an infinitely differentiable function on $\Omega$
satisfying the inequalities
\[
c_1\, d(x) \le \delta(x) \le c_2\, d(x),\qquad \big| \partial_x^\alpha \delta(x)\big|
  \le c_\alpha \, d(x)^{1-|\alpha|}
\]
with certain positive constants $c_1,c_2,c_\alpha$. Furthermore,
let $\rho$ and $\kappa$ be positive numbers, and let $\chi$ be an
infinitely differentiable function such that $0\le \chi \le 1$,
$\chi(t)=0$ for $t\le 0$, and $\chi(t)=1$ for $t\ge 1$. We define
the cut-off function $\zeta$ on $\Omega$ by
\[
\zeta(x) = \chi\Big( \kappa\, \log \frac \rho{\delta(x)}\Big).
\]
This function has the following properties.
\begin{itemize}
\item[(i)] $0\le \zeta(x)\le 1$, $\zeta(x)=0$ for $\delta(x)\ge \rho$, $\zeta(x)=1$ for
  $\delta(x)\le \varepsilon \rho$, where $\varepsilon = e^{-1/\kappa}$.
\item[(ii)] $\displaystyle |\nabla \zeta(x)| \le c\, \frac{\kappa}{d(x)}$,\quad
 $\big|\displaystyle \partial_{x_i}\partial_{x_j} \zeta(x)\big| \le c\frac{\kappa}{d(x)^2}$   for $i,j=1,2,3$.
\end{itemize}
By Lemma \ref{bl3}, the vector function $w$ admits the representation $w=\mbox{rot}\, b$
with a vector function $b\in W^{1,6}(\Omega)^3$ satisfying (\ref{1bl3}). We put
\[
v = V + u, \quad \mbox{where} \quad V=\mbox{rot}(\zeta b) = \zeta w + \nabla\zeta \times b.
\]
Then $u$ satisfies the equations
\begin{equation} \label{9ct1}
- \Delta u +  (v \cdot \nabla)\, u +  (u\cdot\nabla)\, V =
   \Delta V -  (V\cdot \nabla)\, V - \nabla p, \quad  \nabla \cdot u =0
\end{equation}
in $\Omega$ and the boundary condition $u|_{\partial\Omega}=0$. Since
\[
\int_\Omega \big( (v\cdot \nabla)\, u\big)\cdot u\, dx =0,
\]
it follows from (\ref{9ct1}) that
\begin{equation} \label{5ct1}
\sum_{j=1}^3 \| \nabla u_j\|^2_{L_2(\Omega)} - \sum_{j=1}^3 \int_\Omega u_j\, V\cdot
  \frac{\partial u}{\partial x_j}\, dx = L(u),
\end{equation}
where
\begin{eqnarray*}
L(u) & = & \int_\Omega \Big( \Delta V - (V\cdot\nabla )V - \nabla p\Big) \cdot u\, dx
  = \sum_{j=1}^3 \int_\Omega \Big( -\nabla V_j \cdot \nabla u_j
  + V_j\, V\cdot \frac{\partial u}{\partial x_j} \Big)\, dx \\
& = & - \int_\Omega \Big( w\cdot u\, \Delta\zeta + 2w\cdot ( \nabla\zeta\cdot \nabla)\, u+
  q\, u \cdot \nabla\zeta \Big)\, dx - \sum_{j=1}^3 \int_\Omega \nabla (\nabla \zeta \times b)_j \cdot
  \nabla u_j \, dx \\
&& + \  \sum_{j=1}^3 \int_\Omega V_j\, V\cdot \frac{\partial u}{\partial x_j} \, dx
\end{eqnarray*}
(here $(\nabla \zeta \times b)_j$ denotes the $j$th component of the vector $\nabla \zeta \times b$).
We estimate the functional $L(u)$. Using the inequalities
\[
|\nabla \zeta| \le c\, \frac{\kappa}{\varepsilon \rho} \, ,\qquad
  \big|d\, \Delta \zeta\big| \le c\, \frac{\kappa}{\varepsilon \rho} \, ,
\]
\[
\int_\Omega d(x)^{-2}\, \big| u(x)\big|^2\, dx \le c\, \int_\Omega \big| \nabla u(x)\big|^2\, dx
\]
(the last follows from Hardy's inequality) and (\ref{a3}), we obtain
\[
\Big| \int_\Omega w \cdot u\, \Delta\zeta\, dx\Big| \le \| w\|_{L_\infty(\Omega)} \,
  \| d\Delta \zeta\|_{L_2(\Omega)}\, \| d^{-1}u\|_{L_2(\Omega)}  \le  c\, \frac{\kappa}{\varepsilon\rho}\,
  \| \phi \|_{L_\infty(\partial\Omega)}\,   \sum_{j=1}^3 \| \nabla u_j \|_{L_2(\Omega)}
\]
and
\[
\Big| \int_\Omega w\cdot ( \nabla\zeta\cdot \nabla)\, u\, dx\Big| \le \| w\|_{L_\infty(\Omega)}
  \, \| \nabla\zeta\|_{L_2(\Omega)}\, \sum_{j=1}^3 \| \nabla u_j \|_{L_2(\Omega)}
\le  c\, \frac{\kappa}{\varepsilon\rho}\, \| \phi \|_{L_\infty(\partial\Omega)}\,
  \sum_{j=1}^3 \| \nabla u_j \|_{L_2(\Omega)} \, .
\]
Analogously by (\ref{2bl1}) and (\ref{1bl3}),
\[
\Big| \int_\Omega  q\, u \cdot \nabla\zeta \, dx\Big| \le \| qd\|_{L_\infty(\Omega)} \,
  \| d^{-1}u\|_{L_2(\Omega)}\, \| \nabla \zeta\|_{L_2(\Omega)} \le  c\, \frac{\kappa}{\varepsilon\rho}\,
  \| \phi \|_{L_\infty(\partial\Omega)}\,   \sum_{j=1}^3 \| \nabla u_j \|_{L_2(\Omega)} \, ,
\]
\begin{eqnarray*}
\Big| \int_\Omega \nabla (\nabla \zeta \times b)_j \cdot \nabla u_j \, dx \Big| & \le & c\,
  \Big( \| \nabla\zeta\|_{L_\infty(\Omega)}\, \| \nabla b\|_{L_2(\Omega)} + \sum_{i,k}
  \| \frac{\partial^2\zeta}{\partial x_i \partial x_k}\|_{L_\infty(\Omega)} \, \| b\|_{L_2(\Omega)}\Big)\,
  \| \nabla u_j\|_{L_2(\Omega)}\\
& \le & c\, \frac{\kappa}{\varepsilon^2\rho^2}\,  \| \phi \|_{L_\infty(\partial\Omega)}\,
  \| \nabla u_j \|_{L_2(\Omega)}\, ,
\end{eqnarray*}
and
\begin{eqnarray*}
\Big| \int_\Omega V_j\, V\cdot \frac{\partial u}{\partial x_j}\, dx\Big| & \le & \| V \|^2_{L_4(\Omega)}
  \, \| \partial_{x_j} u\|_{L_2(\Omega)}
\le 2\, \Big( \| \zeta w\|^2_{L_4(\Omega)} + \| \nabla\zeta\times b\|^2_{L_4(\Omega)}\Big)\,
  \| \partial_{x_j} u\|_{L_2(\Omega} \\
& \le & c\, \Big( 1+ \frac{\kappa^2}{\varepsilon^2\rho^2}\Big) \, \| \phi\|^2_{L_2(\Omega)}\,
  \| \partial_{x_j} u\|_{L_2(\Omega)}\, .
\end{eqnarray*}
Thus,
\begin{equation} \label{6ct1}
\big| L(u)\big| \le C_1\, \Big( \frac{\kappa}{\varepsilon^2 \rho^2}\, \| \phi\|_{L_\infty(\partial\Omega)}
  + \big( 1+ \frac{\kappa^2}{\varepsilon^2\rho^2}\big)\, \| \phi\|^2_{L_\infty(\partial\Omega)}\Big)
  \, \| \nabla u\|_{L_2(\Omega)}\, ,
\end{equation}
where $C_1$ is a constant independent of $\rho$ and $\kappa$. Furthermore,
\begin{eqnarray*}
&& \Big| \sum_{j=1}^3 \int_\Omega u_j\, V \cdot \frac{\partial u}{\partial x_j}\, dx \Big|  =
  \Big|  \sum_{j=1}^3 \int_\Omega u_j (\zeta w + \nabla\zeta\times b)\cdot \frac{\partial u}{\partial x_j}
   \, dx \Big| \\
&& \le \Big( \| \zeta d\|_{L_\infty(\Omega)}\, \| w\|_{L_\infty(\Omega)} +
  \| d\, \nabla\zeta\|_{L_\infty(\Omega)}\, \| b\|_{L_\infty(\Omega)} \Big) \,
  \sum_{j=1}^3 \| d^{-1} u_j\|_{L_2(\Omega)} \,   \| \partial_{x_j} u\|_{L_2(\Omega)} \\
&& \le C_2\, (\rho+\kappa)\, \| \phi\|_{L_\infty(\Omega)}\, \sum_{j=1}^3 \| \nabla u_j\|_{L_2(\Omega)}\, ,
\end{eqnarray*}
where $C_2$ is independent of $\phi,\rho,\kappa$. The numbers $\rho$ and $\kappa$ can be chosen such that
\[
C_2\, (\rho+\kappa)\,  \| \phi\|_{L_\infty(\partial\Omega)} \le 1/2.
\]
Then it follows from (\ref{5ct1}) and (\ref{6ct1}) that
\[
\sum_{j=1}^3 \| \nabla u_j \|_{L_2(\Omega)} \le 2\, C_1\, \Big( \frac{\kappa}{\varepsilon^2 \rho^2}\,
  \| \phi\|_{L_\infty(\partial\Omega)}
  + \big( 1+ \frac{\kappa^2}{\varepsilon^2\rho^2}\big)\, \| \phi\|^2_{L_\infty(\partial\Omega)}\Big).
\]
By the continuity of the imbedding $W^{1,2}(\Omega) \subset L_6(\Omega)$, the same estimate
(with another constant $C_1$) holds for the norm of $u$ in $L_6(\Omega)^3$. Since
$|\nabla\zeta|\le c\kappa/(\varepsilon\rho)$, we further have
\begin{equation} \label{8ct1}
\| V\|_{L_6(\Omega)} \le \| \zeta w\|_{L_6(\Omega)} + \| \nabla\zeta \times b\|_{L_6(\Omega)}
  \le C_3 \, \big( 1+ \kappa/(\varepsilon \rho)\big)\, \| \phi\|_{L_\infty(\partial\Omega)}
\end{equation}
(see Lemmas \ref{bl1} and \ref{bl3}) and consequently
\[
\| v\|_{L_6(\Omega)} \le \| V\|_{L_6(\Omega)} + \|
u\|_{L_6(\Omega)} \le C_4\, \Big( \big(1+\frac{\kappa}{\varepsilon
  \rho}+\frac{\kappa}{\varepsilon^2 \rho^2}\big)\,   \| \phi\|_{L_\infty(\partial\Omega)}
  + \big( 1+ \frac{\kappa^2}{\varepsilon^2\rho^2}\big)\, \| \phi\|^2_{L_\infty(\partial\Omega)}\Big) .
\]
If we put
\[
\kappa= \rho = \frac{1}{4C_2\, \| \phi\|_{L_\infty(\partial\Omega)}} \quad\mbox{and}\quad
   \varepsilon = e^{-1/\kappa} = e^{-4C_2\| \phi\|_{L_\infty(\partial\Omega)}}\, ,
\]
we obtain
\[
\| v\|_{L_6(\Omega)} \le C_5\, \Big( \| \phi\|_{L_\infty(\partial\Omega)}\,
  e^{4C_2\| \phi\|_{L_\infty(\partial\Omega)}} + \| \phi\|^2_{L_\infty(\partial\Omega)}\,
  e^{8C_2 \| \phi\|_{L_\infty(\partial\Omega)}} \Big) .
\]
This together with (\ref{4ct1}) implies (\ref{a2}) for $\nu=1$. If
$\nu\not=1$, then we consider the vector function $(\nu^{-1}v,\nu^{-2}p)$ instead of $(v,p)$. \hfill $\Box$ \\

\end{document}